\documentclass[aps,prl,preprint,superscriptaddress,floatfix,showpacs,nobibnotes]{revtex4-2}
\usepackage{graphicx,caption}
\usepackage{amssymb}
\usepackage{amsmath} 
\usepackage{amsfonts}
\usepackage{hyperref}
\usepackage{textcomp}
\usepackage{subfig} 
\usepackage{nicefrac,xfrac}
\usepackage{stackengine}
\usepackage{dashrule}
\usepackage{array}
\usepackage{tabularx,comment,footnote}
\usepackage{xtab}
\usepackage{float}
\usepackage[style=base]{caption}
\usepackage{latexsym}
 
\begin{document}

\title{Singular observation of shape and spin hindrance in metastable nuclear decay of $^{197}$Au}
\author{S.K. Tandel \footnote{Corresponding author: sujit.tandel@snu.edu.in, sktandel@gmail.com}}
\affiliation{Department of Physics, School of Natural Sciences, Shiv Nadar Institution of Eminence, Delhi-NCR 201314, India}
\affiliation{Department of Physics, University of Massachusetts Lowell, Lowell, Massachusetts 01854, USA}
\author{D. Rathore}
\affiliation{Atomic and Molecular Physics Division, Bhabha Atomic Research Centre, Mumbai 400085, India}
\author{M. Hemalatha}
\affiliation{Department of Physics, University of Mumbai, Mumbai 400098, India}
\author{Subhrajit Sahoo}
\affiliation{Department of Physics, Indian Institute of Technology Roorkee, Roorkee 247667, India}
\author{P.C. Srivastava}
\affiliation{Department of Physics, Indian Institute of Technology Roorkee, Roorkee 247667, India}
\author{P. Chowdhury} 
\affiliation{Department of Physics, University of Massachusetts Lowell, Lowell, Massachusetts 01854, USA}
\author{R.V.F. Janssens}
\affiliation{Department of Physics and Astronomy, University of North Carolina at Chapel Hill, Chapel Hill, North Carolina 27599, USA}
\affiliation{Triangle Universities Nuclear Laboratory, Duke University, Durham, North Carolina 27708, USA}
\author{F.G. Kondev}
\affiliation{Argonne National Laboratory, Argonne, Illinois 60439, USA}
\author{M.P. Carpenter} 
\affiliation{Argonne National Laboratory, Argonne, Illinois 60439, USA}
\author{T. Lauritsen}
\affiliation{Argonne National Laboratory, Argonne, Illinois 60439, USA}
\author{D. Seweryniak}
\affiliation{Argonne National Laboratory, Argonne, Illinois 60439, USA}

\date{\today}

\begin{abstract}
A {\it I}$^{\pi}$ = (29/2$^{+}$) metastable state with {\it T}$_{1/2}$ $ > $ 10 s at 
{\it E}$_{x}$ = 2102 keV has been identified in the stable nucleus, $^{197}$Au. 
Its decay exhibits the unusual coexistence of a strongly hindered $E2$ transition and a competing 
$E4$ branch with comparable intensity, indicating the simultaneous presence of shape- and spin-related 
hindrance mechanisms. Such a decay pattern has not previously been identified in a long-lived nuclear 
state and serves as a sensitive probe of the interplay between shell structure, configuration mixing, 
and collective degrees of freedom. This result establishes an important benchmark for modern 
nuclear-structure models, for which a unified description of all observables remains challenging.

\end{abstract}

\maketitle

Nuclear excited states are ordinarily short-lived, with lifetimes significantly
shorter than a nanosecond. However, a small fraction have a metastable character 
and are referred to as nuclear isomers. Their transition probabilities are much lower, and 
are associated with a hindrance arising from substantial changes in either shape, spin or 
spin projection {\it K}. Such states are referred to as shape, spin or {\it K} isomers,
respectively \cite{Walker1999}. The lack of overlap of the isomeric wavefunction
with those of levels at lower energy is central to the realization of such isomers,
thereby elucidating the underlying nature of the quantum many-body, strongly interacting 
nuclear system. Additionally, direct insight can be gained into the mechanisms that hinder
electromagnetic decay. Shape isomers result from the rearrangement of nucleons in the transition from 
initial to final states, which are characterized by associated deformations 
quite different from each other in either magnitude or sign \cite{Polikanov1973,Bouchez2003,Moller2012}.
Since angular momentum is a conserved quantum number, an excited level undergoing 
electromagnetic decay through a transition with a high multipole order corresponds to 
a spin isomer \cite{Dracoulis2016,Kondev2015,Jain2015}.
While a large number of shape and spin isomers have been established across light to
heavy nuclei, each of these are specific to either a deformed or a spherical nucleus, 
and as yet, no metastable states with decay hindered on account of both shape and spin 
differences had been established. 

Shape coexistence has been extensively documented in many nuclei, with clear
experimental signatures in some cases and circumstantial evidence in most others 
\cite{Heyde2011,Garrett2022}. Theoretical predictions support a much larger body 
of nuclei which are expected to demonstrate shape coexistence \cite{Moller2012,Heyde2011,Garrett2022}. Shape isomerism,
though, is realized in a small fraction of these nuclei, when the associated minima 
in the nuclear potential have comparable energies \cite{Bouchez2003,Andreyev2000,Plaza2025,Dembski2025}.
The presence of high-{\it j}, unique-parity subshells, which are interspersed with
low-{\it j} orbitals below every major closed shell with nucleon numbers 50, 82 and 
126, creates conditions favorable for spin isomerism with $M3$, $E4$, $M4$ or
even $E5$ decays having been identified \cite{Dracoulis2016,Kondev2015,Jain2015}.
Generally, spin isomers are associated with longer level lifetimes than shape metastable states.
In transitional nuclei having several valence nucleons with respect to closed shells, 
it is possible to realize conditions suitable for both shape and spin hindrance, 
due to the presence of high-{\it j} orbitals and moderately-deformed shapes. 
Such mixed hindrance mechanisms would provide a sensitive probe of the interplay
between collective and single-particle degrees of freedom.
The stable nucleus, $^{197}$Au ({\it Z} = 79, {\it N} = 118), 
belongs to this category, and, therefore, was the subject of this study.

Two experiments were performed to populate long-lived states in $^{197}$Au. The
target was, in both instances, a thick, 50 mg/cm$^{2}$ Au foil, while $^{209}$Bi
and $^{207}$Pb beams with 1450- and 1430-MeV respective energies were used; the 
beam energies were about $10\%$ above the Coulomb barrier, which optimized 
the population of high-spin levels \cite{Broda1990,Cocks2000}.
Pulsed beams were delivered by the ATLAS accelerator at Argonne National Laboratory (ANL),
and $\gamma $-ray coincidence data were recorded using the Gammasphere array which
consisted of 100 Compton-suppressed high-purity germanium (HPGe) detectors at the time
of the experiments \cite{LeeJanssens}. During the experiments, Pb absorbers were placed
in front of the HPGe detectors to suppress the expected copious amount of target x-rays. 
To search for possible long-lived states, the pulsed beam was incident on the target,
and then deflected away for three specific intervals: (a) 200 $\mu $s on and 800 $\mu $s off,
(b) 1 ms on and 3 ms off, and (c) 2 s on and 8 s off, with the data being collected in the
beam-off periods. Additional data were collected in a $\approx $ 1 $\mu $s duration, 
wherein successive beam pulses were separated by 825 ns. In the $\approx $ 1 $\mu $s 
interval, three- and higher-fold coincidence data were recorded, while in the beam-off periods 
of 800 $\mu $s and higher, the minimum fold was two. Some of the histograms
used for the data analysis were: (a) two- and three-dimensional symmetric,
$\gamma $-energy histograms for establishing the excited level structures, and
(b) time-gated, triple-$\gamma $ energy coincidence histograms to determine long half-lives.
Additional information can be found in earlier publications of this group and elsewhere
\cite{Tandel2015,2019Ro12,2020Wa24,2021Su02,2022Bo04,2022Wa20,Tandel2024,Tandel2025,Saket2026,Radford1995}.

A new metastable state in $^{197}$Au, with {\it T}$_{1/2}$ $ > $ 10 s, has been 
established at {\it E}$_{x}$ = 2102 keV, as displayed in Figure 1. This state
deexcites through a 266.6-keV transition to the 1836-keV level with 
{\it I}$^{\pi }$ = (21/2$^{+}$). Another unobserved decay branch of 5.5 keV to the
(25/2$^{+}$) state at {\it E}$_{x}$ = 2097 keV is evident through the observation
of the 261.1-keV $\gamma $ ray in the coincidence spectra of Figure 2. The
half-life of the {\it E}$_{x}$ = 1836-keV state has been newly measured from this
work to be {\it T}$_{1/2}$ = 8.8(6) ns by using the centroid-shift method \cite{Andrejtscheff1982}, 
as illustrated in Figure 3. The level at {\it E}$_{x}$ = 2532 keV with
{\it T}$_{1/2}$ = 150(5) ns reported earlier \cite{Wheldon2006}
is confirmed from the present data. It was not possible to unambiguously 
ascertain the multipolarity of the 435.6-keV $\gamma $ ray deexciting this state. It should 
be noted that transitions with energies 261.1, 357.7, 429.2, 639.7 and 1069.1 keV 
had been established in earlier work \cite{Wheldon2006}. The 266.6-keV $\gamma $ ray was found
to be quite weak in the 1-$\mu $s data, however, in the longer beam-off periods,
it was clearly visible in the data from both experiments. The spectra in Figure 2
correspond to three-fold coincidence events between $\gamma $ rays in the 8-s pulsing
period. The summed integral three-fold coincidence 
counts of all $\gamma $ rays in the deexcitation path of the {\it E}$_{x}$ = 2102-keV
level are displayed in Figure 3(a); these have been obtained from spectra with successively
increasing time bins with 0.5-s time intervals. The regular increase in counts
up to time {\it t} = 8 s indicates a half-life longer than the 8-s beam-off period, with
{\it T}$_{1/2}$ $ > $ 10 s implied from the trend in Figure 3(a).
An inspection of the data collected when the beam was switched off suggests
a half-life longer than several minutes; however, a precise value could not be
determined, therefore, a {\it T}$_{1/2}$ $ > $ 10 s lower limit has been adopted.

The 261.1-keV transition was firmly established to have $E2$ multipolarity,
from the coincidence spectra in the 1-$\mu $s data, based on intensity balance 
considerations between the transitions feeding and deexciting the 1836- and 2097-keV 
levels. Therefore, with {\it I}$^{\pi }$ = (21/2$^{+}$) for the
{\it E}$_{x}$ = 1836-keV level, the spin-parity of the 2097-keV state is
assigned as {\it I}$^{\pi }$ = (25/2$^{+}$). It may be noted that
the 2097-keV level, which is deexcited by the 261.1-keV $\gamma $ ray, 
is fed by a 435.6-keV transition from the decay of a 150(5) ns isomeric state 
at {\it E}$_{x}$ = 2532 keV, as shown in Figure 1 \cite{Wheldon2006}.
Using similar considerations in 
the 8-s pulsing data, the $\gamma $-ray intensity of the 266.6-keV transition
is consistent with an $E4$ character; other multipolarities are excluded, 
as illustrated in Table I. The total intensity ({\it I}$_{t}$) has been calculated
from the measured $\gamma $-ray intensity ({\it I}$_{\gamma }$) using total internal 
conversion coefficients ($\alpha $$_{t}$) from the BrIcc tabulation \cite{Kibedi2008},
and the relation {\it I}$_{t}$ = {\it I}$_{\gamma }$ (1 + $\alpha $$_{t}$). 
For the 266.6-keV transition, {\it I}$_{\gamma }$ = 16(1) gives {\it I}$_{t}$ = 102(6) 
for an $E4$ assignment. This agrees closely with {\it I}$_{t}$ of the subsequent 
357.7- and 639.7-keV transitions, which are 104(4) and 100(4), respectively. 
The corresponding {\it I}$_{t}$ for hypothetical $M2$, $E3$, and $M3$ assignments are 
46(3), 30(2), and 124(8), respectively, inconsistent with the observed intensity balance. 
Multipole orders higher than four, for which the transitions would have much larger
$\alpha $$_{t}$ values, are categorically ruled out. Thus, the $E4$ assignment for
the 266.6-keV line is highly favored by intensity balance considerations. The $M3$ 
assignment is much less likely; further, it would imply $M1$ character for the unobserved 
5.5-keV transition, a conclusion difficult to reconcile with known physical considerations, 
and the non-observation of such a highly-hindered $M1$ transition across the nuclear chart. 
Since the 261.1-keV $\gamma $ ray is visible
in the 8-s pulsing data, the presence of an unobserved 5.5-keV $E2$ transition
is inferred. Thus, {\it I}$^{\pi }$ = (29/2$^{+}$) is proposed for the metastable 
state at {\it E}$_{x}$ = 2102 keV, which is deexcited by both a 5.5-keV $E2$ transition
and a 266.6-keV $E4$ one. 
It is pertinent to note that there is only one other instance of an $E4$ isomeric transition 
in this mass region, {\it viz}., the 129.1-keV $\gamma $ ray deexciting the 
{\it I}$^{\pi }$ = 9$^{-}$ state with {\it T}$_{1/2}$ = 3.54(2) h in $^{202}$Pb \cite{Kondev2024}.

The lower limit on the half-life of the {\it I}$^{\pi }$ = (29/2$^{+}$) state
allows for the determination of upper bounds for the transition
probabilities associated with the 5.5-keV $E2$ and 266.6-keV $E4$ $\gamma $ rays. 
As displayed in Table II, the reduced $E2$ transition probability, 
$B(E2)$ $ < $ $6.2 \times 10^{-9}$ Weisskopf units (W.u.) is indicative of a
highly-suppressed $E2$ transition. The extent of this suppression is comparable
with that observed for classic shape isomers in actinide nuclei, {\it e.g.},
10$^{-7}$ W.u. for the $E2$ decay of the 2558-keV, 0$^{+}$ state in $^{238}$U \cite{Steinmayer1992}. 
In contrast, $B(E4)$ $ < $ 107 W.u. is inferred. A future measurement of 
{\it T}$_{1/2}$(29/2$^{+}$) would affirm whether the $B(E4)$ probability is
enhanced, suppressed or close to the single-particle estimate.

The observation of a highly-suppressed $E2$ transition and an 
$E4$ $\gamma $ ray as decay branches from the same metastable state is without parallel in the
nuclear chart. The only two other instances of $E2$ and $E4$ decays from the
same isomer have been found for the {\it I}$^{\pi }$ = 10$^{+}$ state in $^{54}$Fe, 
with {\it T}$_{1/2}$ = 364(7) ns \cite{Dong2014}, and the 
{\it I}$^{\pi }$ = 19$^{+}$ state in $^{96}$Ag, with {\it T}$_{1/2}$ = 136(20) ns \cite{Boutachkov2011}.
However, unlike in the $^{197}$Au case, $B(E2)$ = 1.69(4) W.u. and $B(E4)$ = 0.79(8) W.u. probabilities
were inferred in $^{54}$Fe, and $B(E2)$ = 4.7(10) W.u. and $B(E4)$ = 0.9(6) W.u.
were deduced in $^{96}$Ag \cite{Dong2014,Boutachkov2011}, indicating the single-particle character of 
these four transitions. No $B(E2)$ suppression is evident in either of the cases. 
In $^{54}$Fe and $^{96}$Ag, the $E2$ decay branches have energies of 146 and 98 keV,
whereas the $E4$ decays proceed through 3578- and 4265-keV transitions, respectively. Thus, the high energy
of the $E4$ transitions enables the competition with the $E2$ $\gamma $ rays, unlike the situation
in $^{197}$Au. 

The origin of the unusual decay pattern in $^{197}$Au may be understood in terms of the simultaneous 
presence of shape- and spin-related hindrance mechanisms. Strongly hindered $E2$ transitions are commonly 
associated with changes in intrinsic deformation between the initial and final states, whereas $E4$ transitions 
with near-single-particle strengths are observed in near-spherical nuclei below major shell closures. 
These two phenomena are generally 
encountered in structurally distinct regions of the nuclear chart. In $^{197}$Au, however, the observation of 
comparable $E2$ and $E4$ decay branches from the same $I^\pi = 29/2^+$ metastable state, together with the 
pronounced suppression of the $E2$ strength, indicates the coexistence of shape and spin hindrance within a 
single decay sequence. To investigate these complementary structural aspects, calculations were performed using 
both the Ultimate Cranker code based on the modified oscillator (Nilsson) potential, which describes deformed 
quasiparticle configurations \cite{Bengtsson1989}, and the KSHELL framework with the KHHE interaction, 
which provides a microscopic description of near-spherical states \cite{Warburton1991,Shimizu2019,Yuan2022}.

The nucleus $^{197}$Au had been reported to be characterized by a moderate oblate deformation in its 
ground state ($\epsilon $$_{2}$ = 0.12) in a recent global tabulation \cite{Moller2016}. This is supported
in the present work by the UC calculations, which also predict deformation parameters
$\epsilon $$_{2}$ = 0.13 and a triaxiality parameter, $\gamma $ = -73$^{o}$
for the first {\it I}$^{\pi}$ = 25/2$^{+}$ level; {\it i.e.}, suggestive of moderate, near-oblate deformation,
as depicted in Figure 4. The {\it E}$_{x}$ = 2097-keV level is, therefore, associated with
this shape. The UC calculations indicate similar oblate deformations for the bands built
on the 11/2$^{-}$ and 21/2$^{+}$ isomeric levels displayed in Figure 1. The shell-model
description is in qualitative agreement, with respective $B(E2: 15/2^{-} \rightarrow 11/2^{-})$ = 25.4 W.u.,
and $B(E2: 25/2^{+} \rightarrow 21/2^{+})$ = 21.1 W.u., values indicative of the collective nature
of these transitions. In contrast, a $B(E2: 29/2^{+} \rightarrow 25/2^{+})$ = 0.6 W.u. value is
calculated, demonstrating an absence of collective enhancement and, thus, the single-particle
character of the {\it I}$^{\pi }$ = 29/2$^{+}$ state. 
The shell-model predictions of the excitation energies of the 21/2$^{+}$, 25/2$^{+}$
and 29/2$^{+}$ states are within 100-200 keV of the experimental
values (Table II). The 25/2$^{+}$ energy from experiment and UC calculations exhibit similar
agreement. Furthermore, empirical calculations have been performed following a prescription 
given in earlier work of this group \cite{2020Wa24,2022Bo04}, where experimental 1-quasiparticle energies, 
pair-gap energies from a 5-point mass formula \cite{Moller1992},
and residual interactions derived from isomeric levels with relatively pure configurations 
in this mass region have been used.
These calculations indicate that the {\it I}$^{\pi }$ = 29/2$^{+}$ level associated with a
$\pi ${\it h}$_{11/2}^{-1}$ $\otimes $ $\nu $({\it i}$_{13/2}^{-1}$, {\it f}$_{5/2}$)
configuration should lie at a rather low energy, {\it E}$_{x}$ = 1911 keV, within 200 keV of 
the experimental value. 
Although the shell-model calculations reproduce the dominant configurations of the states involved, 
they do not explicitly incorporate deformation changes that may contribute to the observed large 
suppression of the $E2$ transition strength in the decay of the {\it I}$^{\pi }$ = 29/2$^{+}$ state. 
Any discrepancy between the experimental and calculated $B(E2)$ values may therefore provide insight 
into shape-related effects beyond the present model space. While a direct comparison of the
experimental and calculated transition probabilities in the decay of the 21/2$^{+}$ level
is possible, and reasonable agreement is seen, the limiting, rather than a precise value of
the 29/2$^{+}$ half-life precludes a similar juxtaposition.

Though long-lived states have been established in odd-{\it A} isotopes of Ir ({\it Z} = 77),
Au ({\it Z} = 79) and Tl ({\it Z} = 81), their decays are distinctly different 
from those observed in $^{197}$Au.
In $^{191,193}$Ir, {\it I}$^{\pi }$ = 31/2$^{+}$ levels associated with triaxial deformation 
and half-lives in the $\mu $s-s range, and deexciting transitions of $M2$, $E3$ or $M3$ 
character, have been established \cite{2012Dr01}. In lighter Au isotopes, {\it viz.}, 
$^{189,191,193,195}$Au, 31/2$^{+}$ and 31/2$^{-}$ levels with half-lives in the ns-$\mu $s
range have been found. These decay by $E2$ transitions in almost all instances
\cite{2007Ok01,2013Dr01}. In $^{203,205}$Tl, $\mu $s isomers with $E2$ and $E3$ decays and predominantly
single-particle character have been identified \cite{2022Bo04,Wrzesinski2003}. Finally, the neutron-deficient $^{188,190}$Pb
isotopes exhibit isomers with spherical and oblate configurations and ns-$\mu $s half-lives.
These levels decay via $E1$, $E2$ or enhanced $E3$ 
($\pi ${\it i}$_{13/2}$ $\rightarrow $ $\pi ${\it f}$_{7/2}$) transitions and considerable
state-mixing is evident in these cases \cite{1998Dr01,Dracoulis2004}.

The coexistence of pronounced shape- and spin-related hindrance in $^{197}$Au is highly unusual 
and has broader implications for the understanding of electromagnetic decay in quantum 
systems, where strongly suppressed transition probabilities are typically associated with a single 
dominant mechanism. In contrast, the decay of the $^{197}$Au isomer appears to involve the simultaneous 
action of two distinct hindrance mechanisms. The simultaneous presence of shape- and spin-related hindrance 
presents a significant challenge for current theoretical approaches, which typically describe either 
collective deformation effects or detailed single-particle configurations, but not both within a unified 
framework. Further, the observation is particularly significant because $^{197}$Au 
lies in the vicinity of the doubly magic nucleus $^{208}$Pb, a benchmark region for nuclear-structure studies 
in which shell closures and single-particle configurations are generally well established. The present 
results demonstrate that structural complexity persists even near this closed-shell region and that 
metastable decay cannot always be understood within a purely single-particle framework. The 
long-lived isomer in $^{197}$Au therefore provides a sensitive probe of the interplay between shell 
structure, configuration mixing, and collective degrees of freedom. These findings provide new 
insight into the persistence of shape-related effects near closed shells and establish an important 
benchmark for modern nuclear-structure models and beyond-mean-field approaches.

In summary, the $I^\pi = 29/2^+$ state in $^{197}$Au with $T_{1/2} > 10$ s exhibits an unusual 
coexistence of shape- and spin-related hindrance in the decay of a metastable nuclear state. The 
observation of a strongly suppressed $E2$ transition competing with a comparable $E4$ branch indicates 
the simultaneous action of collective and single-particle mechanisms within the same decay sequence. 
These results suggest that metastable decay in the vicinity of the doubly magic nucleus $^{208}$Pb may involve 
a more intricate interplay between deformation, shell structure, and configuration mixing than previously recognized, 
with a unified description of all observed properties being challenging. Additional measurements, including 
long-timescale decay studies and mass spectrometry, are planned at ANL to further constrain the structure 
of the $29/2^+$ isomer in $^{197}$Au and establish an exacting yardstick for nuclear theory.

\vskip 12pt
{\bf Acknowledgements}
\vskip 12pt
The authors would like to thank J. Meng, P. Guo and F. Y. He for useful discussions.
S.K.T. acknowledges support from the Shiv Nadar Foundation (SNF). 
This work is supported by the U.S. Department of Energy, Office of Science, Office of 
Nuclear Physics, under award numbers DE-FG02-94ER40848 (UML), DE-FG02-97ER41041 (UNC) 
and DE-FG02-97ER41033 (TUNL), and contract number DE-AC02-06CH11357 (ANL).
The research described here utilized resources of the ATLAS facility at ANL, 
which is a DOE Office of Science user facility.

\bibliographystyle{amsplain}

\begin{table}[t]
\caption{Gamma-ray energies and and total intensities of transitions in $^{197}$Au
observed in the decay of the {\it T}$_{1/2}$ $ > $ 10-s state. The assigned
multipolarities and theoretical conversion coefficients \cite{Kibedi2008} are also provided.}
\label{tab:table1}
\begin{ruledtabular}
\begin{tabular}{ccccc}
	$E_\gamma$ (keV) & Mult. & $\alpha_{tot}^{theo}$ & $I_\gamma$ & $I_t$ \\
\hline

(5.5)   & (E2)  & 1.761$\times10^6$ & 4.0(6)$\times10^{-6}$    & 7(1)   \\
261.1   & E2    & 0.156               & 6(1)                   & 7(1)   \\
266.6   & {\bf E4} & 5.426            & 16(1)                  & {\bf 102(6)} \\
        & M2    & 1.886               &                        & 46(3)   \\
        & E3    & 0.864               &                        & 30(2)   \\
        & M3    & 6.767               &                        & 124(8) \\
357.7   & E2    & 0.062               & 98(4)                  & 104(4)  \\
429.2   & E1    & 0.012               & 100(5)                 & 101(5)  \\
639.7   & E2    & 0.015               & 99(4)                  & 100(4)  \\
1069.1  & E3    & 0.011               & 2(1)                   & 2(1)    \\

\end{tabular}
\end{ruledtabular}
\end{table}

\clearpage

\begin{table*}[t]
\caption{Experimental excitation energies and transition probabilities
for the decay of metastable states in $^{197}$Au determined from
this work along with the results of shell-model calculations using the
KHHE interaction. The $\gamma $-ray energies, multipolarities,
and the spin-parities and half-lives of the levels are indicated.
See text for additional details.}
\label{table:table2}
\begin{center}
\begin{tabular*}{\textwidth}{@{\extracolsep{\fill}} c c c c c c c c}
\hline\hline
\multicolumn{2}{c}{$E_x$ (keV)} & $I^\pi$ & $E_{\gamma}$ (keV) & Mult. & $T_{1/2}^{\text{level}}$ & \multicolumn{2}{c}{$B(E\lambda)$ (W.u.)} \\
\cline{1-2} \cline{7-8}
Exp. & Theo. & & & & & Exp. & Theo. \\
\hline
	1836 & 1936 & $21/2^+$ & 429.2  & $E1$ & 8.8(6) ns & $2.8(1) \times 10^{-7}$ & 0 \\
	&      &          & 1069.1 & $E3$ &           & $0.72(12)$  & $0.33$ \\
        2097 & 2040 & $25/2^+$ & 261.1  & $E2$ & - & - & 21.1 \\
	2102 & 2296 & $29/2^+$ & 5.5    & $E2$ & $> 10$ s & $< 6.2 \times 10^{-9}$ & 0.6 \\
        &      &          & 266.6  & $E4$ &            & $< 107$  &  $3.0 \times 10^{-4}$ \\
\hline
\end{tabular*}
\end{center}
\end{table*}

\clearpage

\begin{figure}
\begin{center}
\includegraphics[scale=0.5]{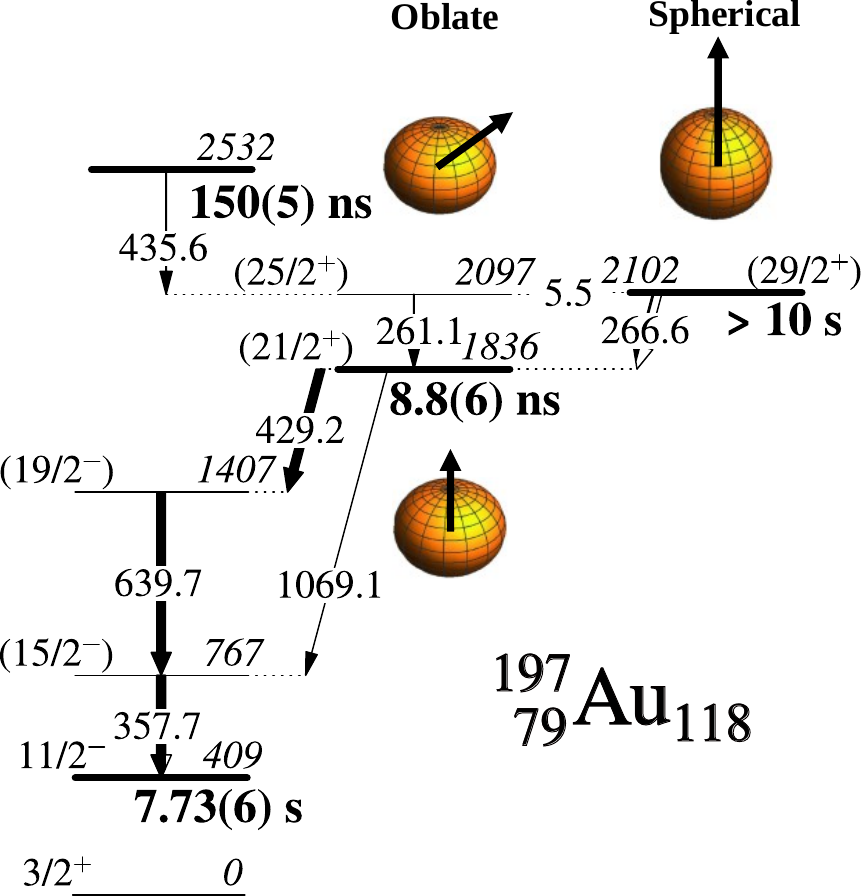}
\caption{Metastable states in $^{197}$Au: the {\it E}$_{x}$ = 2102-keV state
with {\it T}$_{1/2}$ $ > $ 10 s is newly established, while 
{\it T}$_{1/2}$ = 8.8(6) ns for the {\it I}$^{\pi }$ = (21/2$^{+}$) level is measured from this work.} 
\label{fig:fig1}
\end{center}
\end{figure}

\begin{figure}
\begin{center}
\includegraphics[scale=0.36]{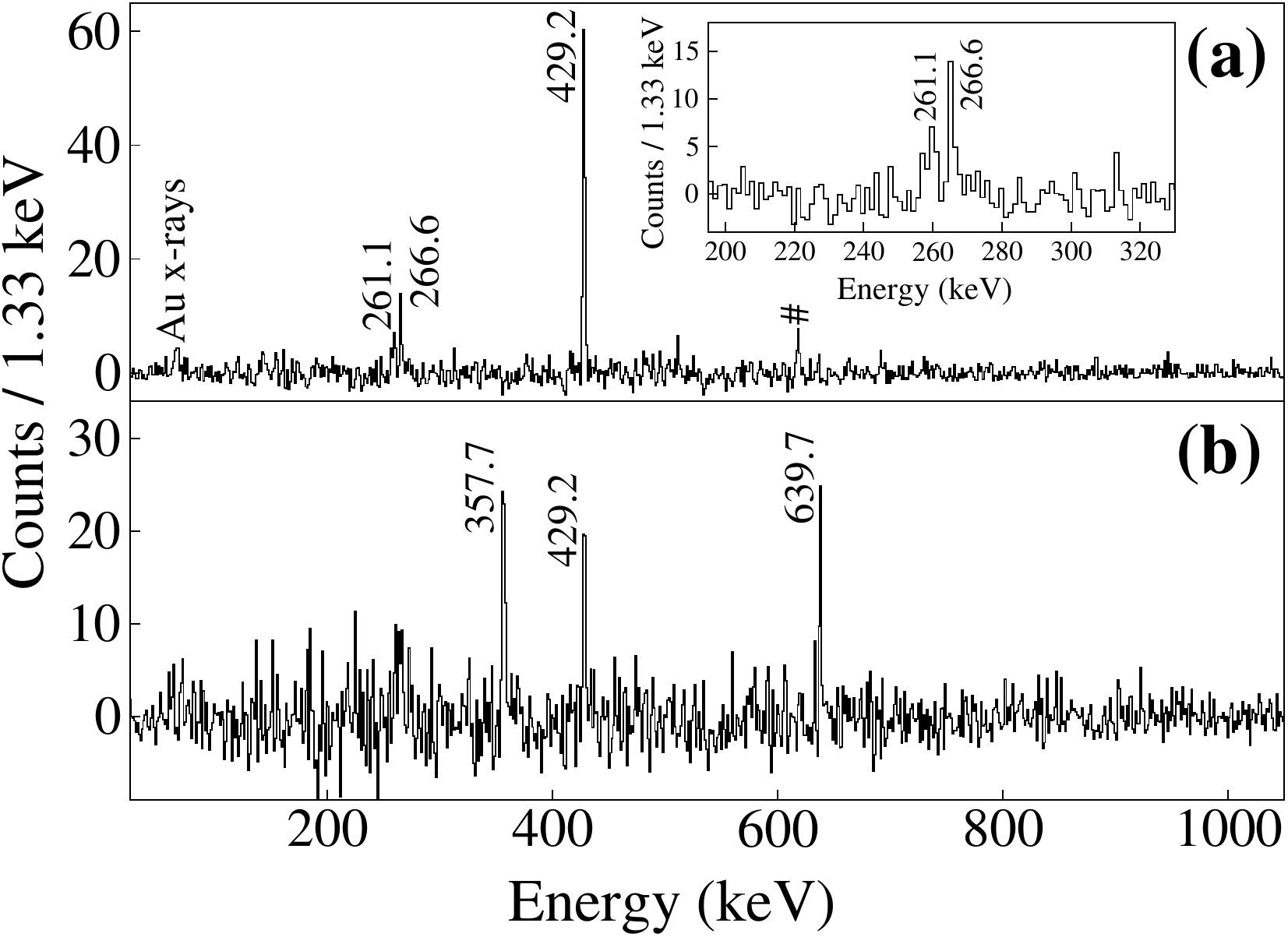}
\caption{Three-fold coincidence spectra illustrating $\gamma $ rays observed
in the decay of the {\it T}$_{1/2}$ $ > $ 10 s state in $^{197}$Au.
The spectrum in panel (a) has been obtained with gates set on the 357.7-
and 639.7-keV transitions, while the one in panel (b) is from the sum of double gates
on the 357.7, 429.2 and 639.7 keV $\gamma $ rays on the one hand, with the 266.6-keV one
deexciting the 2102-keV isomer on the other. The hash mark corresponds to a contaminant 
transition.} 
\label{fig:fig2}
\end{center}
\end{figure}

\begin{figure}
\begin{center}
\includegraphics[scale=0.36]{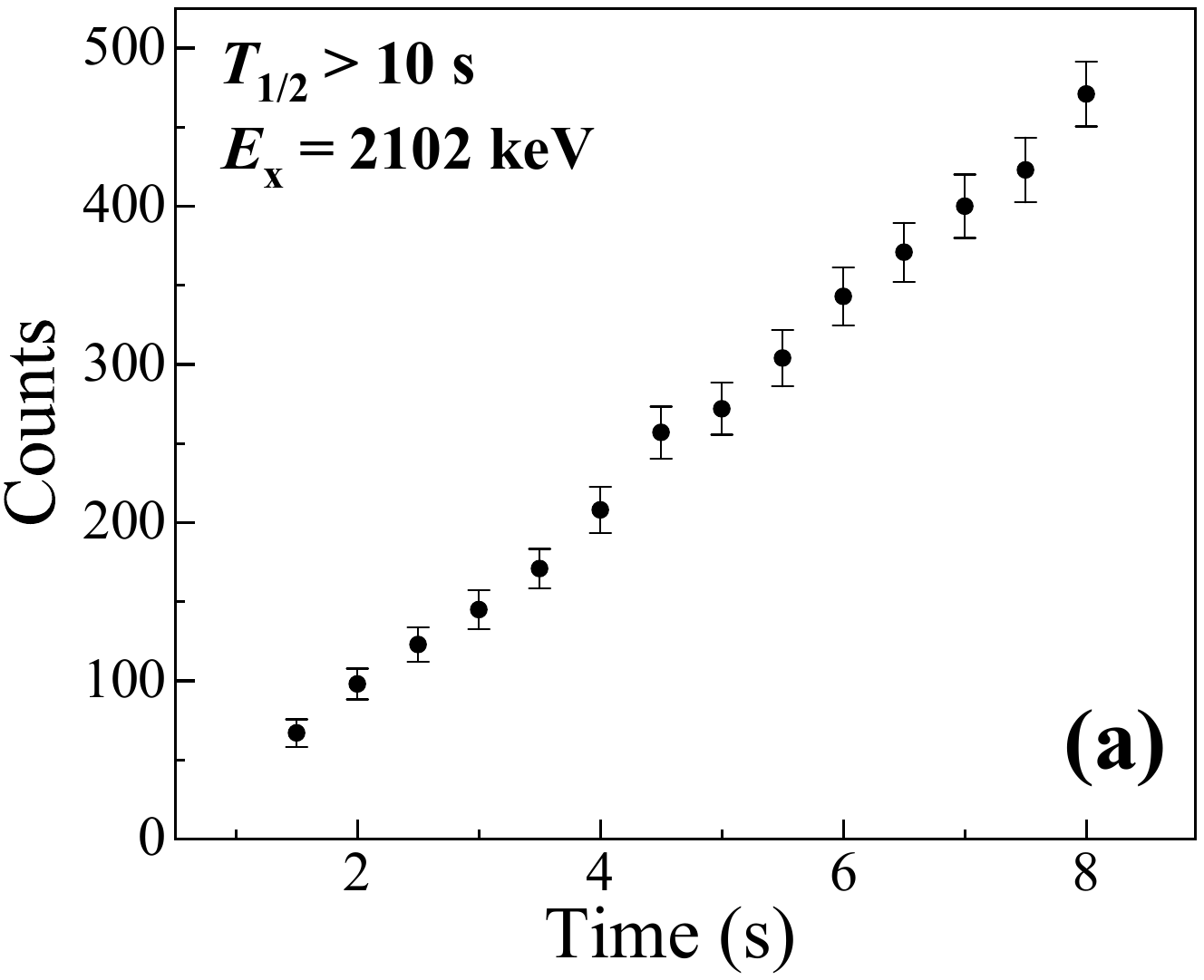}
\includegraphics[scale=0.4]{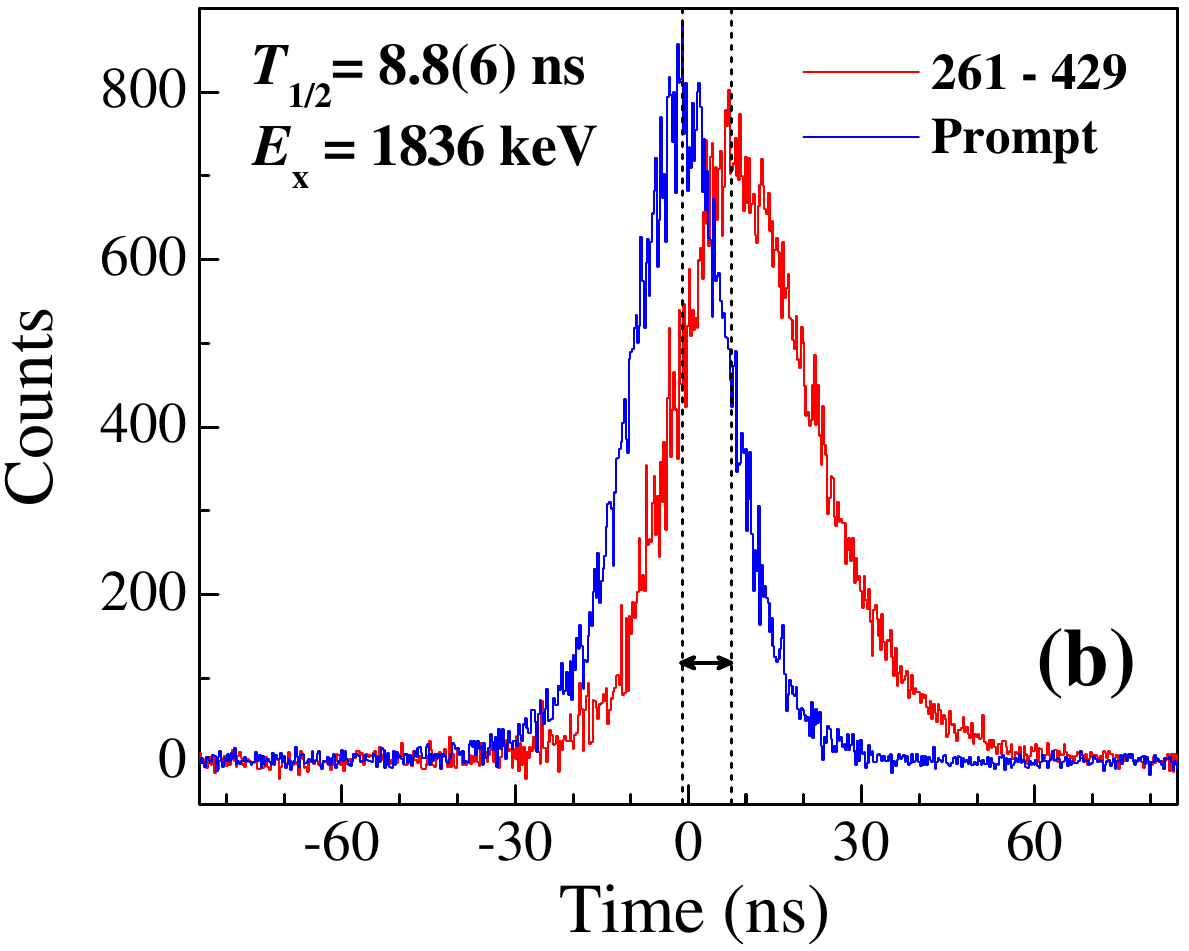}
\caption{(a) Summed integral counts of $\gamma $ rays from the deexcitation of the 
{\it E}$_{x}$ = 2102-keV state, and (b) Half-life of the {\it E}$_{x}$ = 1836-keV
level in $^{197}$Au determined using the centroid-shift method. Details are in the text.}
\label{fig:fig3}
\end{center}
\end{figure}

\begin{figure}
\begin{center}
\includegraphics[scale=0.5]{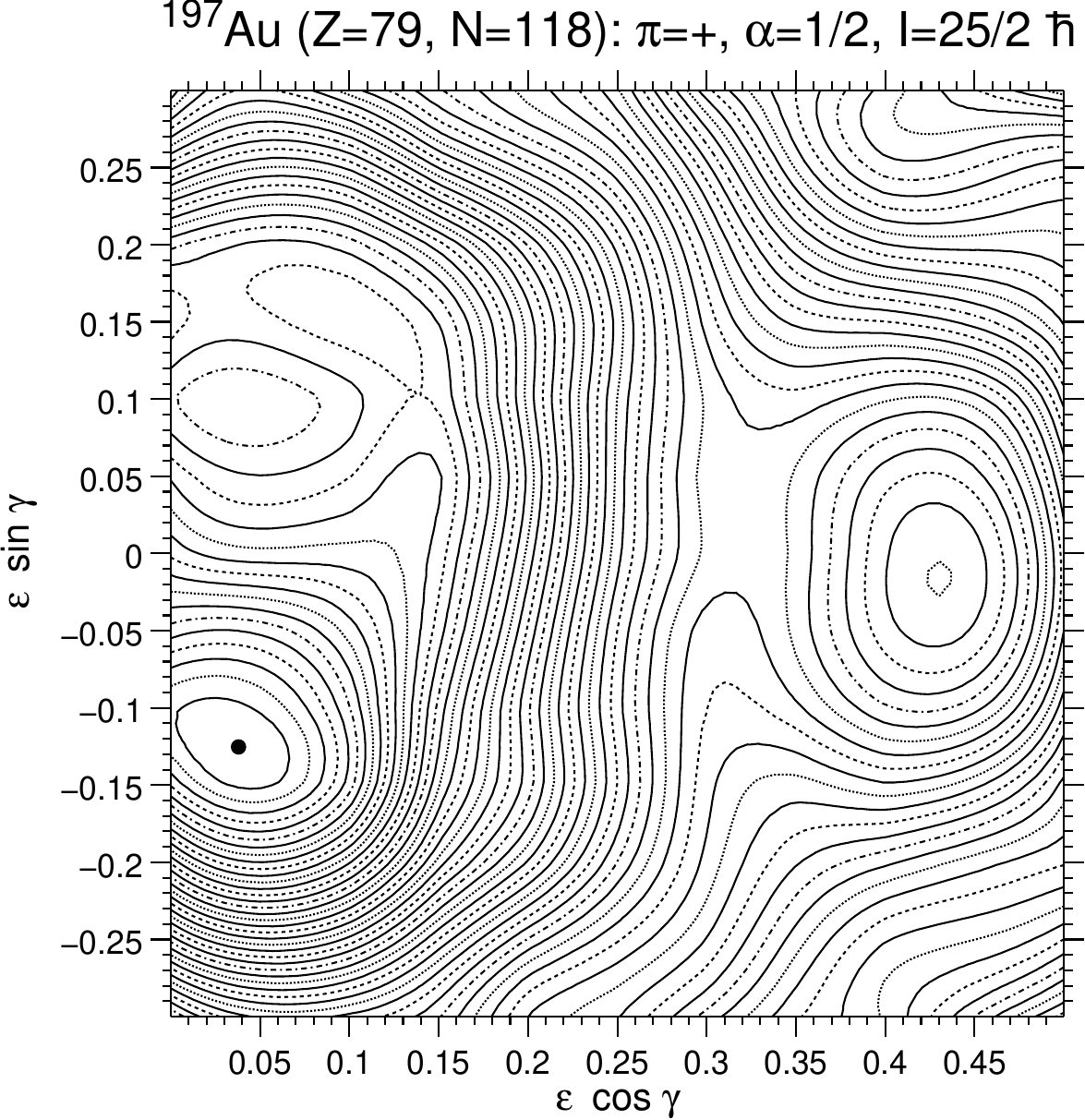}
\caption{Total energy surface for the {\it I}$^{\pi }$ = 25/2$^{+}$ level in
$^{197}$Au calculated using the Ultimate Cranker code \cite{Bengtsson1989}. 
The energy minimum is located at $\epsilon $$_{2}$ = 0.13, $\gamma $ = -73\textdegree.}
\label{fig:fig4}
\end{center}
\end{figure}

\end{document}